\documentclass[aps,prl,twocolumn,floats,showpacs,superscriptaddress]{revtex4-1}

\usepackage{graphicx}
\usepackage{dcolumn}
\usepackage{bm}
\usepackage{amssymb}
\usepackage{braket}
\usepackage{amsmath}
\usepackage{tabularx}
\usepackage{pdfpages}
\makeatletter
\newcommand*{\balancecolsandclearpage}{%
  \close@column@grid
  \clearpage
  \twocolumngrid
}
\makeatother

\begin{document}

\title{Landau levels in strained optical lattices}
\author{Binbin Tian}
\affiliation{University of Pittsburgh, PA 15260, USA}
\author{Manuel Endres}
\affiliation{Department of Physics, Harvard University, Cambridge, MA 02138,
USA}
\affiliation{Institute for Quantum Information and Matter, Department of Physics, California Institute of Technology, Pasadena, CA 91125, USA}
\author{David Pekker}
\affiliation{University of Pittsburgh, PA 15260, USA}

\date{\today}

\begin{abstract}
We propose a hexagonal optical lattice system with spatial variations in the hopping matrix elements. Just like in the valley Hall effect in strained Graphene, for atoms near the Dirac points the variations in the hopping matrix elements can be described by a pseudo-magnetic field and result in the formation of Landau levels. We show that the pseudo-magnetic field leads to measurable experimental signatures in momentum resolved Bragg spectroscopy, Bloch oscillations, cyclotron motion, and quantization of in-situ densities. Our proposal can be realized by a slight modification of existing experiments. In contrast to previous methods, pseudo-magnetic fields are realized in a completely static system avoiding common heating effects and therefore opening the door to studying interaction effects in Landau levels with cold atoms.
\end{abstract}

\maketitle

The Lorentz force, which acts on charged particles moving in a magnetic field, results in a number of fundamental phenomena in condensed matter systems including the Hall effect in metals, Abrikosov lattices in superconductors, and the integer and fractional quantum Hall effects in ultra-pure two-dimensional electron gases. While phenomena, such as the quantized conductance plateaus of the integer and fractional quantum Hall effects have been both observed experimentally and described theoretically~\cite{Doucot2005}, many properties, such as the non-abelian nature of excitations in the fractional quantum Hall effects~\cite{Moore1991, Papic2011, Mong2013}, remain subjects of active research.

These problems are difficult -- they involve strongly interacting systems that are resistant to conventional theoretical and numerical tools. The current theoretical state of the art involves comparisons of trial wave functions and numerical calculations on small systems using exact diagonalization and the density matrix renormalization group (DMRG)~\cite{Papic2011,Mong2013}. Ultra cold atom experiments offer an alternative route, in which, potentially, the interplay of gauge fields, band structure, interactions, and disorder can be studied by engineering and controlling these effects independently~\cite{Jaksch2005}. Moreover, by engineering these properties one could generate novel phases that have not yet been observed in condensed matter systems~\cite{Ghaemi2012,Bitan2014}.

Various groups have recently experimentally demonstrated `synthetic gauge fields' --  methods for driving neutral atoms using laser beams in such a way that they behave as if they were charged particles moving in a magnetic field~\cite{lin2009synthetic, Aidelsburger:2011, miyake2013realizing, Jotzu2014, Goldman2014}. A number of effects, such as Abrikosov lattice formation~\cite{lin2009bose}, Hall deflection~\cite{leblanc2012observation, Jotzu2014, Aidelsburger2014}, and chiral currents~\cite{Atala2014}, have been observed. However, an important limitation of these methods, that use either periodic lattice modulations or Raman transitions, seems to be significant heating of the atom clouds. For optical lattice experiments, this limits the timescale in which experiments can be performed to several tens of milliseconds~\cite{supplement}, in contrast to experiments in static lattices which allow for several hundreds of milliseconds~\cite{bloch2008many}. Further, experiments so far have been performed with noninteracting or weakly interacting atoms. An extension of synthetic gauge fields experiments to the strongly interacting regime would require low heating rates in combination with low initial temperatures.

\begin{figure*}
\includegraphics[width=\textwidth]{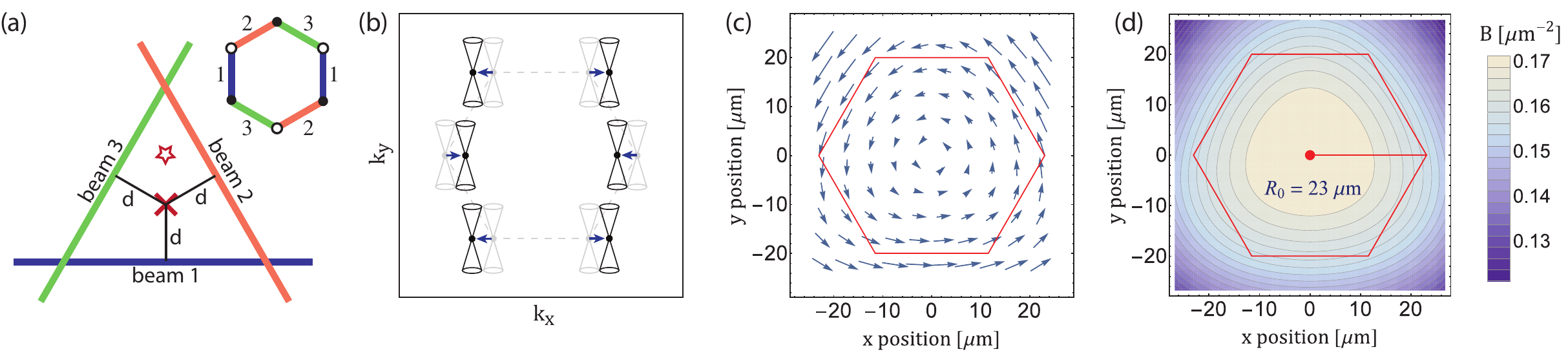}
\caption{(a) Schematic of the setup: the optical lattice is produced by three Gaussian laser beams intersecting at $120^\circ$ with offset $d$ -- lattice orientation is depicted in the upper right. The cross marks the center of the harmonic trap, and the star marks the position associated with the displaced Dirac cones depicted in (b). (b) Schematic of the displacement of the Dirac cones in momentum space associated with stretching type 1 bond. First Brillouin zone is indicated with the dashed line. (c) Pseudo-vector potential $\vec{A}$ as a function of position. (d) Pseudo-magnetic field $\vec{B}=\nabla \times \vec{A}$ as a function of position. The hexagon in (c) and (d) marks the sample area with 80\% uniformity in the pseudo-magnetic field.
}
\label{fig:schematic}
\end{figure*}

In this Letter, we propose an alternative method for generating synthetic magnetic fields in ultracold atom systems that relies on a completely static optical lattice, and leads to the formation of relativistic Landau levels. Inspired by pseudo-magnetic fields observed in strained graphene~\cite{Guinea2010,levy2010strain}, molecular graphene~\cite{Gomes2012}, and photonic systems~\cite{Rechtsman2013}, we propose a method for generating spatially varying hopping matrix elements in a hexagonal optical lattice. Starting from a standard configuration of three Gaussian laser beams intersecting at $120^\circ$, our method relies on simply displacing the beams [Fig.~\ref{fig:schematic}a]. We show that pronounced Landau levels are generated close to the original Dirac points corresponding to almost homogeneous magnetic fields in real space with opposite sign at the nonequivalent Dirac points. This leads to a host of observable phenomena, such as the valley Hall effect, quantization of in-situ densities, Landau-Zener effects in Bloch oscillations, and the emergence of gap structure in Bragg spectroscopy. Since our method relies on fully static optical lattices, it forms an attractive starting point for investigating interaction effects in Landau levels~\cite{Herbut2008, Abanin2012,Ghaemi2012,Gopalakrishnan2012, Roy2013, Roy2014, Roy2014b,Bitan2014} with ultracold atoms.

{\it Pseudo magnetic fields in optical lattices --\/} Consider an optical lattice that is created by the intersection of three blue-detuned laser beams at $120^\circ$ angles~\cite{soltan2011multi,Duca2015}. The resulting honeycomb potential has the form
\begin{align}
V(\vec{r})&=\left| \sum_{m=1}^3 \sqrt{I_m(\vec{r})} e^{-i \vec{k}_m \cdot \vec{r}}\right|^2
\end{align}
where $\vec{k}_m$ is the wave vector of the $m$-th laser beam and $I_m(\vec{r})$ is it's position dependent intensity that accounts for the Gaussian nature of the beams. We consider the case that the lattice is sufficiently deep so that the tight binding model is applicable. If all three beams have identical intensities the resulting band structure (in the lowest band) is identical to that of graphene with two degeneracy points in the first Brillouin zone described by the Dirac equation. Changing the intensity of one of the beams corresponds to applying a uniform strain to graphene~\cite{alba2013simulating}, which can be captured by modifying the hopping matrix elements in the tight binding model and results in the shift of the Dirac cones in the Brillouin zone [see Fig.~\ref{fig:schematic}b]. This shift can be encoded in terms of the Dirac equation by adding a vector gauge field $\vec{A}$:
\begin{align}
H_\text{Dirac}=\hbar v_f \left(-i \partial_\mu + A_\mu\right) \sigma^\mu,
\end{align}
where $v_f$ is the group velocity near the Dirac point, and $\sigma^\mu$ are the Pauli matrices. Allowing for the intensities of all three beams to vary in space, we can obtain a non-uniform $\vec{A}(\vec{r})$.

The tight binding model with non-uniform hopping matrix elements on the honeycomb lattice is~\footnote{Since we are interested in the case $I_m\gtrsim 2 E_R$ we only need to consider nearest neighbor hopping ($t_\text{nnn}=0.02 t_\text{nn}$ at $2 E_R$).}
\begin{align}
H=-\sum_{\langle{i,j}\rangle} t_{ij} a_{i}^\dagger a_{j} + \sum_{i} V_i a_{i}^\dagger a_{i},
\label{eq:Hamiltonian}
\end{align}
where $t_{ij}$ are the hopping matrix elements, $V_i=\frac{1}{2} m_\text{at} \omega_\text{eff}^2 r_i^2$ is the on-site potential that is a combination of the trap potential and the anti-trapping effect of the blue lattice $\omega_\text{eff}^2=\omega_\text{trap}^2-\omega_\text{anti-trap}^2$ with $m_\text{at}$ the atomic mass, and $a^\dagger_i$ and $a_i$ are the creation and annihilation operators. There are three types of hopping matrix elements $t_{ij}$ associated with the three hopping directions in a honeycomb lattice. We label these as $t_{ij}=t_{u}(\vec{r})$, where $\vec{r}=(\vec{r}_i+\vec{r}_j)/2$ and $u$ is the index of the laser beam that is perpendicular to the vector $\vec{r}_{i}-\vec{r}_{j}$ (see Fig.~\ref{fig:schematic}a). Using this notation, and making the approximation that $t_{1}(\vec{r}) \simeq t_{2}(\vec{r}) \simeq t_{3}(\vec{r})$ and all are slowly varying functions of $\vec{r}$, we find
\begin{align}
\left(
\begin{array}{c}
A_x(\vec{r})\\
A_y(\vec{r})
\end{array}\right)
=\pm \frac{\sqrt{3}}{2 \lambda t_0(\vec{r})} \left(
\begin{array}{c}
2 t_3(\vec{r})-t_1(\vec{r})-t_2(\vec{r})\\
\sqrt{3}(t_2(\vec{r})-t_1(\vec{r}))
\end{array}
\right). \label{eq:A}
\end{align}
Here, $\lambda$ is the wavelength of the laser, $t_0(\vec{r})=(t_1(\vec{r})+t_2(\vec{r})+t_3(\vec{r}))/3$, and $\pm$ corresponds to the gauge field at the two distinct Dirac points. We note that this form of the vector gauge is identical to that obtained in strained graphene, but the origin of the variation in the hopping matrix elements is different -- here it is induced by spatial variation of the lattice depth as opposed to strain in graphene~\cite{Suzuura2002, Manes2007}.

We connect the tight binding parameters $t_u(\vec{r})$ and $V(\vec{r})$ to the laser light intensity $I_u(\vec{r})$ at point $\vec{r}$ using a simple analytical model that accurately captures a precise numerical calculation~\cite{walters2013ab}. Our model~\cite{supplement} assumes that the connection is completely local [valid when $I_u(\vec{r})$ varies slowly in space on the length-scale of a unit cell] and that the laser beam intensities are close to isotropic [$I_1(\vec{r})\simeq I_2(\vec{r}) \simeq I_3(\vec{r})$] -- precisely the assumptions required for the validity of the pseudo-magnetic field description.

\begin{figure*}
  \centering
\includegraphics[width=17cm]{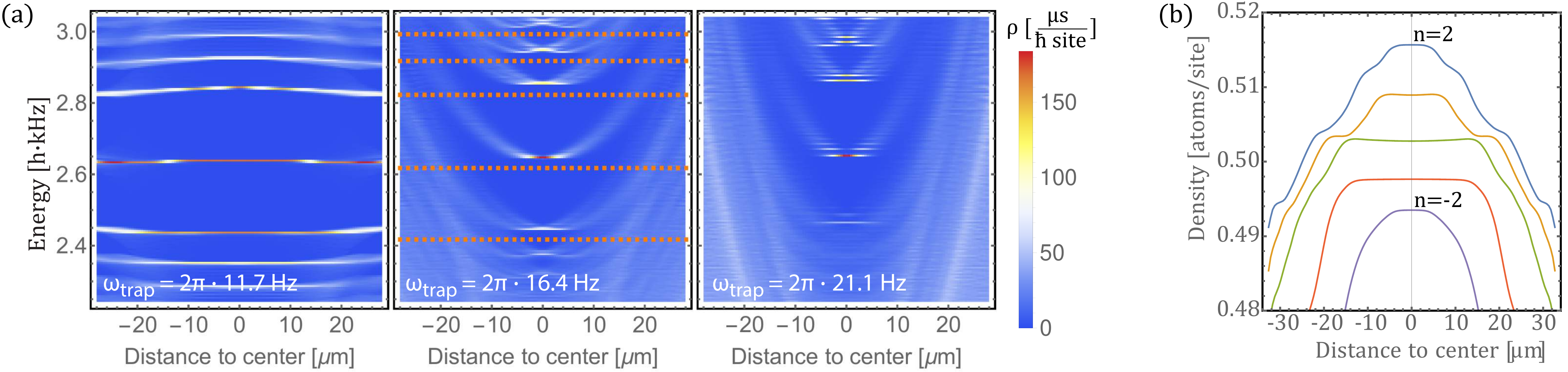}
  \caption{(a) Local Density of States %(LDOS)
  as a function of energy and position in the trap for various trap frequencies. $\omega_\text{trap}=11.7\,(2\pi\,\text{Hz})$ -- trap cancels the anti-trapping potential of lattice beams. $\omega_\text{trap}=16.4\,(2\pi\,\text{Hz})$ -- bending of distinct Landau. $\omega_\text{trap}=21.1\,(2\pi\,\text{Hz})$ -- strong smearing of Landau levels. (b) Density of a fermionic fluid as a function of position in the trap showing incompressible plateaus [corresponding chemical potentials are indicated with dashed lines in (a)].}\label{fig:LDOS}
\end{figure*}

{\it A prescription for a uniform pseudo-magnetic field -- } To introduce spatial variations of beam intensities, we propose to use three Gaussian beams with the same intensity and beam waist and to align them such that the beam axis form an equilateral triangle Fig.~\ref{fig:schematic}a. This choice results in a symmetric gauge field depicted in Fig.~\ref{fig:schematic}c, and a nearly  uniform pseudo-magnetic field depicted in Fig.~\ref{fig:schematic}d. We note that this prescription does not introduce an offset between A and B sub-lattices. We can tune the strength and uniformity of the pseudo-magnetic field by varying the beam waist size $w_0$ and the displacement parameter $d$ [Fig.~\ref{fig:schematic}a]. In order to probe uniform phases of matter, such as the quantized Hall effects, one typically requires a uniform magnetic field over the sample area. We have tabulated the optimal choice of $w_0$ and $d$ for a range of sample sizes $R_0$ that ensure a uniform pseudo-magnetic field~\cite{supplement}.

The maximum pseudo-magnetic field is limited because the description in terms of the Dirac equation with $\vec{A}$ breaks down when the displacement of the Dirac cones becomes comparable to the the size of the Brillouin zone $|\vec{A}|\sim 1/\lambda$. For the symmetric gauge choice $|\vec{A}|$ varies linearly across the sample, and hence the maximal pseudo-magnetic field is inversely proportional to the sample size. From Table~II of the supplement we find $B_\text{max} = \nabla \times \vec{A} \approx \text{2.7}/\lambda R_0$.

{\it Proposed experimental setup --} To make a concrete connection to experiment, we focus on a particular realization similar to the one in Ref.~\cite{Duca2015}, which we will use throughout the remainder of the letter. We consider $^{87}$Rb atoms in an optical lattice with $\lambda = 700\text{nm}$, and hence a recoil energy of $E_R=2\pi^2\hbar^2/(m\lambda^2)\approx h \times 4.685\text{kHz}$. We choose the laser beam intensity such that the potential is $4 E_R$ (per beam, in the beam center) which ensures that the lattice is sufficiently deep to make the tight binding model applicable but sufficiently shallow that the hopping timescales are acceptably fast [$t_{ij}/h \approx 868\text{Hz}$]. Finally, we focus on the case $R_0=23 \mu \text{m}$, which corresponds to a hexagonal sample with $15,000$ sites. For this setup, the maximum uniform pseudo-magnetic field of $B=0.17\mu m^{-2}$ [Fig.~\ref{fig:schematic}d] is obtained by setting $w_0=150 \mu m$ and $d=45 \mu m$.

\begin{figure*}
  \centering
  \includegraphics[width=0.7\textwidth]{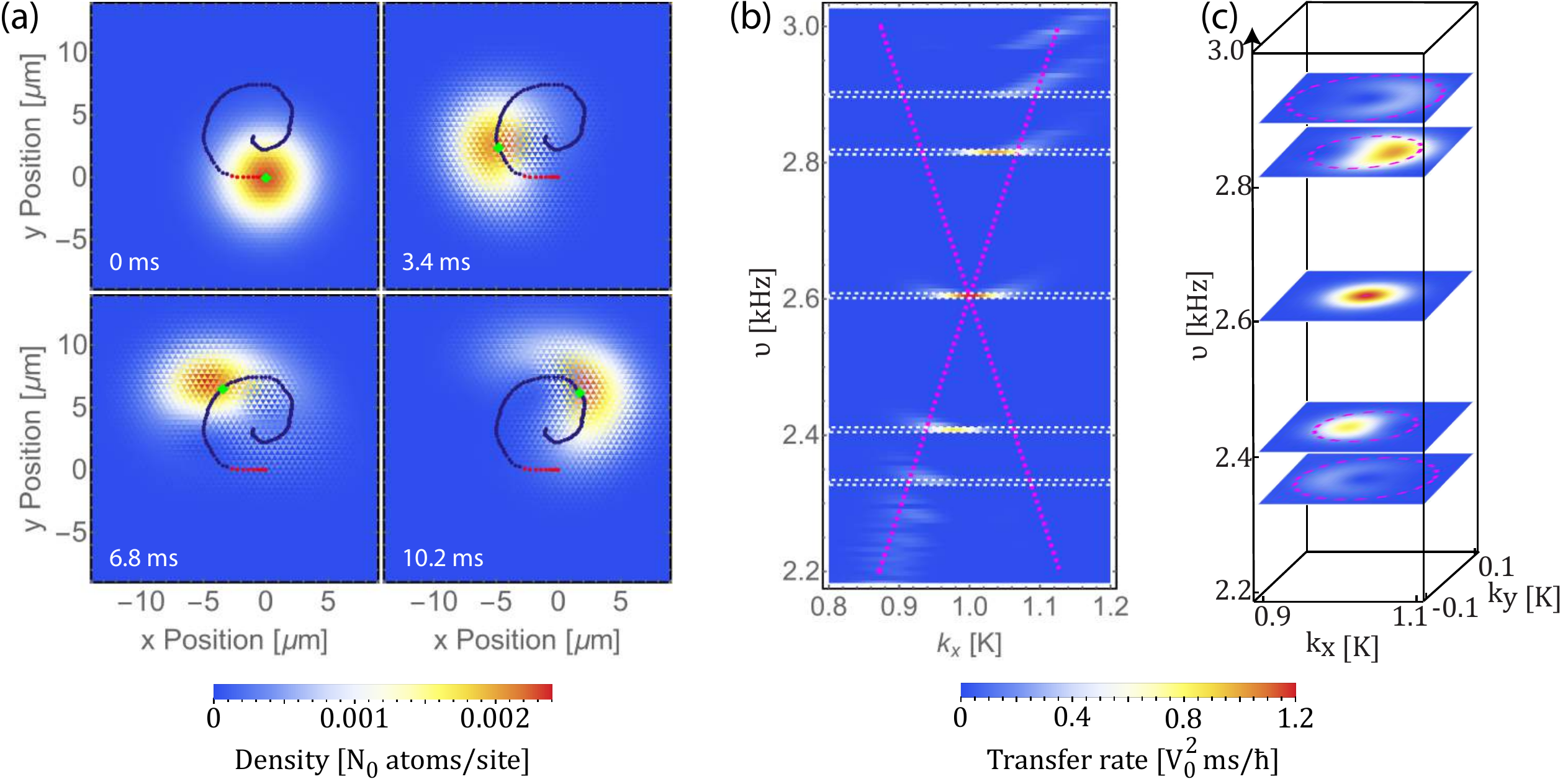}
  \caption{(a) Snap shots of the in-situ density of an atom cloud undergoing cyclotron motion, where $N_0$ is the total number of atoms in the cloud. Trajectory of the center of mass of the cloud is superimposed on top (red dots -- tilt $\alpha/h=0.84 \text{kHz}/\mu\text{m}$, blue -- no tilt, green -- current frame). (b)-(c) Momentum resolved Bragg-spectroscopy: transition rate as a function of frequency $\nu=\omega/2\pi$ and momenta $k_x$ and $k_y$. (b) Slice at fixed $k_y$=0. (c) Slices at fixed $\omega$ (as indicated by the dashed white lines in (b)). The location of the Dirac cone is indicated with dashed magenta line in (b) and (c).}\label{fig:bosonic}
\end{figure*}

{\it Landau levels in a harmonic trap --} The effective trapping frequency $\omega_\text{eff}$ must be large enough to ensure that the atoms are confined, but not so large as to smear out the discrete Landau levels. Without a trapping potential, we expect to see `relativistic' Landau levels at energies $E_n=\text{sign}(n) \hbar v_F \sqrt{2 |n| B}$, where $v_F=\lambda t_{ij}/(\sqrt{3} \hbar)$ and $n$ is an integer~\cite{Berestetskii1971,Novoselov2005}. The corresponding Landau level wave functions have a length scale set by the cyclotron radius $r_c=\sqrt{(2n+1)/B}$. The length scale for the ground state of the trapped lattice system is $\lambda_\text{SHO}=\left[2t_{ij} \lambda^2/(9m_\text{Rb} \omega_\text{eff}^2)\right]^{1/4}$, hence the ground state wave function will be confined if $\lambda_\text{SHO}\lesssim R_0$, which sets a lower bound on $\omega_\text{trap}$. On the other hand, a Landau level will avoid being smeared out if the shift of the trap potential on the scale of $r_c$ is small compared to the spacing to the next Landau level $\frac{1}{2} m_\text{Rb} \omega_\text{eff}^2 r_c^2 \lesssim E_{|n|+1}-E_{|n|}$, which sets an upper bound on $\omega_\text{eff}$. Putting these considerations together, we find a window for observing Landau Levels
\begin{align}
\frac{2\lambda^2 t_{ij}}{9m_\text{Rb} R_0^4}
\lesssim \omega_\text{eff}^2 \lesssim
\frac{7.2 t_{ij}}{m_\text{Rb}  R_0^{3/2} \lambda^{1/2}}\left(\frac{\sqrt{|n+1|}-\sqrt{|n|}}{2|n|+1}\right)\notag,
\end{align}
where in the last expression we have used the relation between $R_0$ and $B$. For our realization $0.2 \text{Hz} \lesssim \omega_\text{eff}/2\pi \lesssim  \{91 \text{Hz}, 12 \text{Hz}\}$ for $n=0,\pm1$.

To visualize the Landau Levels in a trapped system we plot $\rho(E,r)$, the local density of states (LDOS), as a function of position and energy [Fig.~\ref{fig:LDOS}a]. In the absence of a trap, we expect to see sharp peaks in the LDOS as a function of energy that are independent of position. For $|\nabla V(r)| \ll |E_n-E_{n-1}|/r_c$, the LDOS is simply shifted $\rho(E,r) \approx \rho(E-V(r),0)$ while for large $|\nabla V(r)|$, this picture breaks down and the peaks in $\rho(E,r)$ become smeared.

{\it Experimental signatures -- } One of the defining characteristics of the quantum Hall effect is the formation of incompressible plateaus associated with the Landau levels. We propose that these plateaus can be observed with ultra cold fermionic atoms by measuring the atom density as a function of distance away from the center of the trap [Fig.~\ref{fig:LDOS}b]. In principle, it should be possible to observe multiple plateaus in a ``wedding cake" structure. Due to the limitations on experimentally accessible optical lattice sizes we expect to see only one or two plateaus.

Another important limitation of the present generation of experiments with fermions in optical lattices is that the lowest temperatures that one can achieve $T\sim 0.5 t_{ij}$~\cite{Greif2013, Hart2015} are similar to the spacing of the Landau levels that we predict for the smallest samples~\cite{supplement}. An attractive alternative is to use bosonic atoms that can be Bose-Einstein condensed (BEC) in the lattice.

To detect the pseudo-magnetic field we propose observing cyclotron motion of a cloud of condensed atoms. Starting with the cloud in the ground state in the center of the trap, we remove the trap (set $\omega_\text{eff}=0$) and accelerate the cloud towards one of the Dirac points by applying a tilt $H_\text{tilt}=-\alpha \,  \hat{q} \cdot \vec{r}$ ($\hat{q}$ is a unit vector indicating tilt direction, and $\alpha$ is the tilt `force'). When the momenta of the atoms approaches the Dirac point, we remove the tilt and observe the evolution of the cloud in situ. As the atom cloud is moving with a momentum close to the Dirac momentum, it is well described by the Dirac equation, and hence its trajectory is curved by the pseudo-magnetic field [Fig.~\ref{fig:bosonic}a]. The effective Lorentz force changes sign if the cloud is accelerated towards a non-equivalent Dirac point.

In order to directly observe the Landau levels, we propose to use Bragg spectroscopy. In this setup, a BEC that is originally prepared in the ground state $\psi_0$ is transferred to excited states $\psi_i$ using a Bragg spectroscopy setup~\cite{Ernst2010,supplement} described by the perturbation potential $V(r,t)=V_1 \cos(\vec{k}\cdot\vec{r}) \cos(\omega t)$. In Fig.~\ref{fig:bosonic}(b), (c) we plot the transition rate for a range of $\vec{k}$ and $\omega$. As a function of $\omega$, we see clear maxima that correspond to the various Landau levels. Moreover, we can extract the momentum structure of the Landau level wave functions by varying $\vec{k}$ (i.e. performing momentum resolved Bragg spectroscopy). Without the complexity of a Bragg spectroscopy setup, Bloch-Zener spectroscopy~\cite{Kling2010} offers an alternative that can detect the separation between the $n=0$ and $n=\pm1$ Landau levels~\cite{supplement}.

{\it Observing interaction effects--\/}
Pseudo-Landau levels are particularly susceptible to the effects of interactions as the kinetic energy of the atoms within a single pseudo-Landau level is quenched. Previous theoretical investigations showed that the interplay of short and long range interactions in strained hexagonal lattices can drive the formation of a number of exotic phases~\cite{Herbut2008, Abanin2012,Ghaemi2012,Gopalakrishnan2012, Roy2013, Roy2014, Roy2014b,Bitan2014}. Specifically, at 1/2-filling of the zeroth Landau level, long range interactions prefer the formation of the quantum anomalous Hall effect (that spontaneously breaks time reversal symmetry)~\cite{Herbut2008, Abanin2012,Roy2013, Roy2014, Roy2014b} while short range interactions prefer the formation of charge-density-wave phases~\cite{Gopalakrishnan2012, Roy2014b}. Detailed numerical analysis of the 1/2 filling case with on-site interactions alone suggests the formation of an exotic mixture of ferro- and anti-ferromagnetism~\cite{Bitan2014}. At 2/3 filling of the zeroth Landau level, tuning the ratio of long and short range interactions is predicted to drive the formation of a triplet superfluid, a fractional topological insulator and the 2/3 fractional quantum Hall effect~\cite{Ghaemi2012}. In comparison to natural graphene, our proposal gives the advantages of (1) better control over the strain patterns and (2) control over both short~\cite{Jaksch2005} and long-rang interactions (using methods based on Rydberg atoms~\cite{macri2014rydberg}, dipolar atoms~\cite{de2013nonequilibrium}, or dipolar molecules~\cite{hazzard2014many}) and (3) control over filling factors and (4) the availability of low disorder potentials, making it particularly promising for realizing these exotic phases.

{\it Outlook--\/}
We have proposed a scheme for generating a pseudo-magnetic field in ultra cold atom systems that relies on spatial variations of the hopping matrix in analogy to the case of strained graphene. The typical timescales for conventional synthetic gauge field experiments in optical lattices is several tens of milliseconds~\cite{supplement} mostly limited by drive induced heating. Our approach could extend this to several hundreds of milliseconds as typical in static lattices~\cite{bloch2008many}, thus improving timescales for investigating the interplay of pseudo-magnetic fields and interactions. Generally, our work could establish a new, very explicit link between solid state physics and ultracold quantum gases connecting ongoing experimental work in strained graphene and novel, strained optical lattices.

\begin{acknowledgments}
{\it Acknowledgements\/}
It is our pleasure to thank Chandra Varma for his suggestion to consider `strained graphene' optical lattices. We also thank Ulrich Schneider, Tracy Li, and Immanuel Bloch for useful discussions, especially in regards to experimental realizations. D.~P. acknowledges support from the Pittsburgh Quantum Institute, M.~E. acknowledges support from the Harvard Quantum Optics Center and the Caltech Institute for Quantum Information and Matter.
\end{acknowledgments}

\bibliography{mybibnew}

\begin{thebibliography}{44}%
\makeatletter
\providecommand \@ifxundefined [1]{%
 \@ifx{#1\undefined}
}%
\providecommand \@ifnum [1]{%
 \ifnum #1\expandafter \@firstoftwo
 \else \expandafter \@secondoftwo
 \fi
}%
\providecommand \@ifx [1]{%
 \ifx #1\expandafter \@firstoftwo
 \else \expandafter \@secondoftwo
 \fi
}%
\providecommand \natexlab [1]{#1}%
\providecommand \enquote  [1]{``#1''}%
\providecommand \bibnamefont  [1]{#1}%
\providecommand \bibfnamefont [1]{#1}%
\providecommand \citenamefont [1]{#1}%
\providecommand \href@noop [0]{\@secondoftwo}%
\providecommand \href [0]{\begingroup \@sanitize@url \@href}%
\providecommand \@href[1]{\@@startlink{#1}\@@href}%
\providecommand \@@href[1]{\endgroup#1\@@endlink}%
\providecommand \@sanitize@url [0]{\catcode `\\12\catcode `\$12\catcode
  `\&12\catcode `\#12\catcode `\^12\catcode `\_12\catcode `\%12\relax}%
\providecommand \@@startlink[1]{}%
\providecommand \@@endlink[0]{}%
\providecommand \url  [0]{\begingroup\@sanitize@url \@url }%
\providecommand \@url [1]{\endgroup\@href {#1}{\urlprefix }}%
\providecommand \urlprefix  [0]{URL }%
\providecommand \Eprint [0]{\href }%
\providecommand \doibase [0]{http://dx.doi.org/}%
\providecommand \selectlanguage [0]{\@gobble}%
\providecommand \bibinfo  [0]{\@secondoftwo}%
\providecommand \bibfield  [0]{\@secondoftwo}%
\providecommand \translation [1]{[#1]}%
\providecommand \BibitemOpen [0]{}%
\providecommand \bibitemStop [0]{}%
\providecommand \bibitemNoStop [0]{.\EOS\space}%
\providecommand \EOS [0]{\spacefactor3000\relax}%
\providecommand \BibitemShut  [1]{\csname bibitem#1\endcsname}%
\let\auto@bib@innerbib\@empty
%</preamble>
\bibitem [{\citenamefont {Dou{\c c}ot}\ \emph {et~al.}(2005)\citenamefont
  {Dou{\c c}ot}, \citenamefont {Duplantier}, \citenamefont {Pasquier},\ and\
  \citenamefont {Rivasseau}}]{Doucot2005}%
  \BibitemOpen
  \bibinfo {editor} {\bibfnamefont {B.}~\bibnamefont {Dou{\c c}ot}}, \bibinfo
  {editor} {\bibfnamefont {B.}~\bibnamefont {Duplantier}}, \bibinfo {editor}
  {\bibfnamefont {V.}~\bibnamefont {Pasquier}}, \ and\ \bibinfo {editor}
  {\bibfnamefont {V.}~\bibnamefont {Rivasseau}},\ eds.,\ \href@noop {} {\emph
  {\bibinfo {title} {The Quantum Hall Effect: Poincar{\'e} Seminar}}}\
  (\bibinfo  {publisher} {Birkh{\"a}user Verlag, Basel},\ \bibinfo {year}
  {2005})\BibitemShut {NoStop}%
\bibitem [{\citenamefont {Moore}\ and\ \citenamefont {Read}(1991)}]{Moore1991}%
  \BibitemOpen
  \bibfield  {author} {\bibinfo {author} {\bibfnamefont {G.}~\bibnamefont
  {Moore}}\ and\ \bibinfo {author} {\bibfnamefont {N.}~\bibnamefont {Read}},\
  }\href {\doibase http://dx.doi.org/10.1016/0550-3213(91)90407-O} {\bibfield
  {journal} {\bibinfo  {journal} {Nucl. Phys. B}\ }\textbf {\bibinfo {volume}
  {360}},\ \bibinfo {pages} {362 } (\bibinfo {year} {1991})}\BibitemShut
  {NoStop}%
\bibitem [{\citenamefont {Papi\ifmmode~\acute{c}\else \'{c}\fi{}}\ \emph
  {et~al.}(2011)\citenamefont {Papi\ifmmode~\acute{c}\else \'{c}\fi{}},
  \citenamefont {Bernevig},\ and\ \citenamefont {Regnault}}]{Papic2011}%
  \BibitemOpen
  \bibfield  {author} {\bibinfo {author} {\bibfnamefont {Z.}~\bibnamefont
  {Papi\ifmmode~\acute{c}\else \'{c}\fi{}}}, \bibinfo {author} {\bibfnamefont
  {B.~A.}\ \bibnamefont {Bernevig}}, \ and\ \bibinfo {author} {\bibfnamefont
  {N.}~\bibnamefont {Regnault}},\ }\href {\doibase
  10.1103/PhysRevLett.106.056801} {\bibfield  {journal} {\bibinfo  {journal}
  {Phys. Rev. Lett.}\ }\textbf {\bibinfo {volume} {106}},\ \bibinfo {pages}
  {056801} (\bibinfo {year} {2011})}\BibitemShut {NoStop}%
\bibitem [{\citenamefont {Zaletel}\ \emph {et~al.}(2013)\citenamefont
  {Zaletel}, \citenamefont {Mong},\ and\ \citenamefont {Pollmann}}]{Mong2013}%
  \BibitemOpen
  \bibfield  {author} {\bibinfo {author} {\bibfnamefont {M.~P.}\ \bibnamefont
  {Zaletel}}, \bibinfo {author} {\bibfnamefont {R.~S.~K.}\ \bibnamefont
  {Mong}}, \ and\ \bibinfo {author} {\bibfnamefont {F.}~\bibnamefont
  {Pollmann}},\ }\href {\doibase 10.1103/PhysRevLett.110.236801} {\bibfield
  {journal} {\bibinfo  {journal} {Phys. Rev. Lett.}\ }\textbf {\bibinfo
  {volume} {110}},\ \bibinfo {pages} {236801} (\bibinfo {year}
  {2013})}\BibitemShut {NoStop}%
\bibitem [{\citenamefont {Jaksch}\ and\ \citenamefont
  {Zoller}(2005)}]{Jaksch2005}%
  \BibitemOpen
  \bibfield  {author} {\bibinfo {author} {\bibfnamefont {D.}~\bibnamefont
  {Jaksch}}\ and\ \bibinfo {author} {\bibfnamefont {P.}~\bibnamefont
  {Zoller}},\ }\href {\doibase http://dx.doi.org/10.1016/j.aop.2004.09.010}
  {\bibfield  {journal} {\bibinfo  {journal} {Ann. Phys.}\ }\textbf {\bibinfo
  {volume} {315}},\ \bibinfo {pages} {52 } (\bibinfo {year}
  {2005})}\BibitemShut {NoStop}%
\bibitem [{\citenamefont {Ghaemi}\ \emph {et~al.}(2012)\citenamefont {Ghaemi},
  \citenamefont {Cayssol}, \citenamefont {Sheng},\ and\ \citenamefont
  {Vishwanath}}]{Ghaemi2012}%
  \BibitemOpen
  \bibfield  {author} {\bibinfo {author} {\bibfnamefont {P.}~\bibnamefont
  {Ghaemi}}, \bibinfo {author} {\bibfnamefont {J.}~\bibnamefont {Cayssol}},
  \bibinfo {author} {\bibfnamefont {D.~N.}\ \bibnamefont {Sheng}}, \ and\
  \bibinfo {author} {\bibfnamefont {A.}~\bibnamefont {Vishwanath}},\
  }\href@noop {} {\bibfield  {journal} {\bibinfo  {journal} {Phys. Rev. Lett.}\
  }\textbf {\bibinfo {volume} {108}},\ \bibinfo {pages} {266801} (\bibinfo
  {year} {2012})}\BibitemShut {NoStop}%
\bibitem [{\citenamefont {Roy}\ \emph {et~al.}(2014{\natexlab{a}})\citenamefont
  {Roy}, \citenamefont {Assaad},\ and\ \citenamefont {Herbut}}]{Bitan2014}%
  \BibitemOpen
  \bibfield  {author} {\bibinfo {author} {\bibfnamefont {B.}~\bibnamefont
  {Roy}}, \bibinfo {author} {\bibfnamefont {F.~F.}\ \bibnamefont {Assaad}}, \
  and\ \bibinfo {author} {\bibfnamefont {I.~F.}\ \bibnamefont {Herbut}},\
  }\href {\doibase 10.1103/PhysRevX.4.021042} {\bibfield  {journal} {\bibinfo
  {journal} {Phys. Rev. X}\ }\textbf {\bibinfo {volume} {4}},\ \bibinfo {pages}
  {021042} (\bibinfo {year} {2014}{\natexlab{a}})}\BibitemShut {NoStop}%
\bibitem [{\citenamefont {Lin}\ \emph {et~al.}(2009{\natexlab{a}})\citenamefont
  {Lin}, \citenamefont {Compton}, \citenamefont {Jimenez-Garcia}, \citenamefont
  {Porto},\ and\ \citenamefont {Spielman}}]{lin2009synthetic}%
  \BibitemOpen
  \bibfield  {author} {\bibinfo {author} {\bibfnamefont {Y.-J.}\ \bibnamefont
  {Lin}}, \bibinfo {author} {\bibfnamefont {R.~L.}\ \bibnamefont {Compton}},
  \bibinfo {author} {\bibfnamefont {K.}~\bibnamefont {Jimenez-Garcia}},
  \bibinfo {author} {\bibfnamefont {J.~V.}\ \bibnamefont {Porto}}, \ and\
  \bibinfo {author} {\bibfnamefont {I.~B.}\ \bibnamefont {Spielman}},\
  }\href@noop {} {\bibfield  {journal} {\bibinfo  {journal} {Nature}\ }\textbf
  {\bibinfo {volume} {462}},\ \bibinfo {pages} {628} (\bibinfo {year}
  {2009}{\natexlab{a}})}\BibitemShut {NoStop}%
\bibitem [{\citenamefont {Aidelsburger}\ \emph {et~al.}(2011)\citenamefont
  {Aidelsburger}, \citenamefont {Atala}, \citenamefont {Nascimb\`ene},
  \citenamefont {Trotzky}, \citenamefont {Chen},\ and\ \citenamefont
  {Bloch}}]{Aidelsburger:2011}%
  \BibitemOpen
  \bibfield  {author} {\bibinfo {author} {\bibfnamefont {M.}~\bibnamefont
  {Aidelsburger}}, \bibinfo {author} {\bibfnamefont {M.}~\bibnamefont {Atala}},
  \bibinfo {author} {\bibfnamefont {S.}~\bibnamefont {Nascimb\`ene}}, \bibinfo
  {author} {\bibfnamefont {S.}~\bibnamefont {Trotzky}}, \bibinfo {author}
  {\bibfnamefont {Y.-A.}\ \bibnamefont {Chen}}, \ and\ \bibinfo {author}
  {\bibfnamefont {I.}~\bibnamefont {Bloch}},\ }\href {\doibase
  10.1103/PhysRevLett.107.255301} {\bibfield  {journal} {\bibinfo  {journal}
  {Phys. Rev. Lett.}\ }\textbf {\bibinfo {volume} {107}},\ \bibinfo {pages}
  {255301} (\bibinfo {year} {2011})}\BibitemShut {NoStop}%
\bibitem [{\citenamefont {Miyake}\ \emph {et~al.}(2013)\citenamefont {Miyake},
  \citenamefont {Siviloglou}, \citenamefont {Kennedy}, \citenamefont {Burton},\
  and\ \citenamefont {Ketterle}}]{miyake2013realizing}%
  \BibitemOpen
  \bibfield  {author} {\bibinfo {author} {\bibfnamefont {H.}~\bibnamefont
  {Miyake}}, \bibinfo {author} {\bibfnamefont {G.~A.}\ \bibnamefont
  {Siviloglou}}, \bibinfo {author} {\bibfnamefont {C.~J.}\ \bibnamefont
  {Kennedy}}, \bibinfo {author} {\bibfnamefont {W.~C.}\ \bibnamefont {Burton}},
  \ and\ \bibinfo {author} {\bibfnamefont {W.}~\bibnamefont {Ketterle}},\
  }\href@noop {} {\bibfield  {journal} {\bibinfo  {journal} {Phys. Rev. Lett.}\
  }\textbf {\bibinfo {volume} {111}},\ \bibinfo {pages} {185302} (\bibinfo
  {year} {2013})}\BibitemShut {NoStop}%
\bibitem [{\citenamefont {Jotzu}\ \emph {et~al.}(2014)\citenamefont {Jotzu},
  \citenamefont {Messer}, \citenamefont {Desbuquois}, \citenamefont {Lebrat},
  \citenamefont {Uehlinger}, \citenamefont {Greif},\ and\ \citenamefont
  {Esslinger}}]{Jotzu2014}%
  \BibitemOpen
  \bibfield  {author} {\bibinfo {author} {\bibfnamefont {G.}~\bibnamefont
  {Jotzu}}, \bibinfo {author} {\bibfnamefont {M.}~\bibnamefont {Messer}},
  \bibinfo {author} {\bibfnamefont {R.}~\bibnamefont {Desbuquois}}, \bibinfo
  {author} {\bibfnamefont {M.}~\bibnamefont {Lebrat}}, \bibinfo {author}
  {\bibfnamefont {T.}~\bibnamefont {Uehlinger}}, \bibinfo {author}
  {\bibfnamefont {D.}~\bibnamefont {Greif}}, \ and\ \bibinfo {author}
  {\bibfnamefont {T.}~\bibnamefont {Esslinger}},\ }\href
  {http://dx.doi.org/10.1038/nature13915} {\bibfield  {journal} {\bibinfo
  {journal} {Nature}\ }\textbf {\bibinfo {volume} {515}},\ \bibinfo {pages}
  {237} (\bibinfo {year} {2014})}\BibitemShut {NoStop}%
\bibitem [{\citenamefont {Goldman}\ \emph {et~al.}(2014)\citenamefont
  {Goldman}, \citenamefont {Juzeli{\=u}nas}, \citenamefont {{\"O}hberg},\ and\
  \citenamefont {Spielman}}]{Goldman2014}%
  \BibitemOpen
  \bibfield  {author} {\bibinfo {author} {\bibfnamefont {N.}~\bibnamefont
  {Goldman}}, \bibinfo {author} {\bibfnamefont {G.}~\bibnamefont
  {Juzeli{\=u}nas}}, \bibinfo {author} {\bibfnamefont {P.}~\bibnamefont
  {{\"O}hberg}}, \ and\ \bibinfo {author} {\bibfnamefont {I.~B.}\ \bibnamefont
  {Spielman}},\ }\href {http://stacks.iop.org/0034-4885/77/i=12/a=126401}
  {\bibfield  {journal} {\bibinfo  {journal} {Rep. Prog. Phys.}\ }\textbf
  {\bibinfo {volume} {77}},\ \bibinfo {pages} {126401} (\bibinfo {year}
  {2014})}\BibitemShut {NoStop}%
\bibitem [{\citenamefont {Lin}\ \emph {et~al.}(2009{\natexlab{b}})\citenamefont
  {Lin}, \citenamefont {Compton}, \citenamefont {Perry}, \citenamefont
  {Phillips}, \citenamefont {Porto},\ and\ \citenamefont
  {Spielman}}]{lin2009bose}%
  \BibitemOpen
  \bibfield  {author} {\bibinfo {author} {\bibfnamefont {Y.-J.}\ \bibnamefont
  {Lin}}, \bibinfo {author} {\bibfnamefont {R.~L.}\ \bibnamefont {Compton}},
  \bibinfo {author} {\bibfnamefont {A.~R.}\ \bibnamefont {Perry}}, \bibinfo
  {author} {\bibfnamefont {W.~D.}\ \bibnamefont {Phillips}}, \bibinfo {author}
  {\bibfnamefont {J.~V.}\ \bibnamefont {Porto}}, \ and\ \bibinfo {author}
  {\bibfnamefont {I.~B.}\ \bibnamefont {Spielman}},\ }\href@noop {} {\bibfield
  {journal} {\bibinfo  {journal} {Phys. Rev. Lett.}\ }\textbf {\bibinfo
  {volume} {102}},\ \bibinfo {pages} {130401} (\bibinfo {year}
  {2009}{\natexlab{b}})}\BibitemShut {NoStop}%
\bibitem [{\citenamefont {LeBlanc}\ \emph {et~al.}(2012)\citenamefont
  {LeBlanc}, \citenamefont {Jim{\'e}nez-Garc{\'\i}a}, \citenamefont {Williams},
  \citenamefont {Beeler}, \citenamefont {Perry}, \citenamefont {Phillips},\
  and\ \citenamefont {Spielman}}]{leblanc2012observation}%
  \BibitemOpen
  \bibfield  {author} {\bibinfo {author} {\bibfnamefont {L.~J.}\ \bibnamefont
  {LeBlanc}}, \bibinfo {author} {\bibfnamefont {K.}~\bibnamefont
  {Jim{\'e}nez-Garc{\'\i}a}}, \bibinfo {author} {\bibfnamefont {R.~A.}\
  \bibnamefont {Williams}}, \bibinfo {author} {\bibfnamefont {M.~C.}\
  \bibnamefont {Beeler}}, \bibinfo {author} {\bibfnamefont {A.~R.}\
  \bibnamefont {Perry}}, \bibinfo {author} {\bibfnamefont {W.~D.}\ \bibnamefont
  {Phillips}}, \ and\ \bibinfo {author} {\bibfnamefont {I.~B.}\ \bibnamefont
  {Spielman}},\ }\href@noop {} {\bibfield  {journal} {\bibinfo  {journal}
  {Proc. Natl. Acad. Sci. USA}\ }\textbf {\bibinfo {volume} {109}},\ \bibinfo
  {pages} {10811} (\bibinfo {year} {2012})}\BibitemShut {NoStop}%
\bibitem [{\citenamefont {Aidelsburger}\ \emph {et~al.}(2015)\citenamefont
  {Aidelsburger}, \citenamefont {Lohse}, \citenamefont {Schweizer},
  \citenamefont {Atala}, \citenamefont {Barreiro}, \citenamefont {Nascimbene},
  \citenamefont {Cooper}, \citenamefont {Bloch},\ and\ \citenamefont
  {Goldman}}]{Aidelsburger2014}%
  \BibitemOpen
  \bibfield  {author} {\bibinfo {author} {\bibfnamefont {M.}~\bibnamefont
  {Aidelsburger}}, \bibinfo {author} {\bibfnamefont {M.}~\bibnamefont {Lohse}},
  \bibinfo {author} {\bibfnamefont {C.}~\bibnamefont {Schweizer}}, \bibinfo
  {author} {\bibfnamefont {M.}~\bibnamefont {Atala}}, \bibinfo {author}
  {\bibfnamefont {J.~T.}\ \bibnamefont {Barreiro}}, \bibinfo {author}
  {\bibfnamefont {S.}~\bibnamefont {Nascimbene}}, \bibinfo {author}
  {\bibfnamefont {N.~R.}\ \bibnamefont {Cooper}}, \bibinfo {author}
  {\bibfnamefont {I.}~\bibnamefont {Bloch}}, \ and\ \bibinfo {author}
  {\bibfnamefont {N.}~\bibnamefont {Goldman}},\ }\href@noop {} {\bibfield
  {journal} {\bibinfo  {journal} {Nature Phys.}\ }\textbf {\bibinfo {volume}
  {11}},\ \bibinfo {pages} {162} (\bibinfo {year} {2015})}\BibitemShut
  {NoStop}%
\bibitem [{\citenamefont {Atala}\ \emph {et~al.}(2014)\citenamefont {Atala},
  \citenamefont {Aidelsburger}, \citenamefont {Lohse}, \citenamefont
  {Barreiro}, \citenamefont {Paredes},\ and\ \citenamefont
  {Bloch}}]{Atala2014}%
  \BibitemOpen
  \bibfield  {author} {\bibinfo {author} {\bibfnamefont {M.}~\bibnamefont
  {Atala}}, \bibinfo {author} {\bibfnamefont {M.}~\bibnamefont {Aidelsburger}},
  \bibinfo {author} {\bibfnamefont {M.}~\bibnamefont {Lohse}}, \bibinfo
  {author} {\bibfnamefont {J.~T.}\ \bibnamefont {Barreiro}}, \bibinfo {author}
  {\bibfnamefont {B.}~\bibnamefont {Paredes}}, \ and\ \bibinfo {author}
  {\bibfnamefont {I.}~\bibnamefont {Bloch}},\ }\href@noop {} {\bibfield
  {journal} {\bibinfo  {journal} {Nature Phys.}\ }\textbf {\bibinfo {volume}
  {10}},\ \bibinfo {pages} {588} (\bibinfo {year} {2014})}\BibitemShut
  {NoStop}%
\bibitem [{sup()}]{supplement}%
  \BibitemOpen
  \href@noop {} {}\bibinfo {note} {See the supplement}\BibitemShut {NoStop}%
\bibitem [{\citenamefont {Bloch}\ \emph {et~al.}(2008)\citenamefont {Bloch},
  \citenamefont {Dalibard},\ and\ \citenamefont {Zwerger}}]{bloch2008many}%
  \BibitemOpen
  \bibfield  {author} {\bibinfo {author} {\bibfnamefont {I.}~\bibnamefont
  {Bloch}}, \bibinfo {author} {\bibfnamefont {J.}~\bibnamefont {Dalibard}}, \
  and\ \bibinfo {author} {\bibfnamefont {W.}~\bibnamefont {Zwerger}},\
  }\href@noop {} {\bibfield  {journal} {\bibinfo  {journal} {Rev. Mod. Phys.}\
  }\textbf {\bibinfo {volume} {80}},\ \bibinfo {pages} {885} (\bibinfo {year}
  {2008})}\BibitemShut {NoStop}%
\bibitem [{\citenamefont {Guinea}\ \emph {et~al.}(2010)\citenamefont {Guinea},
  \citenamefont {Katsnelson},\ and\ \citenamefont {Geim}}]{Guinea2010}%
  \BibitemOpen
  \bibfield  {author} {\bibinfo {author} {\bibfnamefont {F.}~\bibnamefont
  {Guinea}}, \bibinfo {author} {\bibfnamefont {M.~I.}\ \bibnamefont
  {Katsnelson}}, \ and\ \bibinfo {author} {\bibfnamefont {A.~K.}\ \bibnamefont
  {Geim}},\ }\href {http://dx.doi.org/10.1038/nphys1420} {\bibfield  {journal}
  {\bibinfo  {journal} {Nature Phys.}\ }\textbf {\bibinfo {volume} {6}},\
  \bibinfo {pages} {30} (\bibinfo {year} {2010})}\BibitemShut {NoStop}%
\bibitem [{\citenamefont {Levy}\ \emph {et~al.}(2010)\citenamefont {Levy},
  \citenamefont {Burke}, \citenamefont {Meaker}, \citenamefont {Panlasigui},
  \citenamefont {Zettl}, \citenamefont {Guinea}, \citenamefont {Neto},\ and\
  \citenamefont {Crommie}}]{levy2010strain}%
  \BibitemOpen
  \bibfield  {author} {\bibinfo {author} {\bibfnamefont {N.}~\bibnamefont
  {Levy}}, \bibinfo {author} {\bibfnamefont {S.}~\bibnamefont {Burke}},
  \bibinfo {author} {\bibfnamefont {K.}~\bibnamefont {Meaker}}, \bibinfo
  {author} {\bibfnamefont {M.}~\bibnamefont {Panlasigui}}, \bibinfo {author}
  {\bibfnamefont {A.}~\bibnamefont {Zettl}}, \bibinfo {author} {\bibfnamefont
  {F.}~\bibnamefont {Guinea}}, \bibinfo {author} {\bibfnamefont {A.~C.}\
  \bibnamefont {Neto}}, \ and\ \bibinfo {author} {\bibfnamefont
  {M.}~\bibnamefont {Crommie}},\ }\href@noop {} {\bibfield  {journal} {\bibinfo
   {journal} {Science}\ }\textbf {\bibinfo {volume} {329}},\ \bibinfo {pages}
  {544} (\bibinfo {year} {2010})}\BibitemShut {NoStop}%
\bibitem [{\citenamefont {Gomes}\ \emph {et~al.}(2012)\citenamefont {Gomes},
  \citenamefont {Mar}, \citenamefont {Ko}, \citenamefont {Guinea},\ and\
  \citenamefont {Manoharan}}]{Gomes2012}%
  \BibitemOpen
  \bibfield  {author} {\bibinfo {author} {\bibfnamefont {K.~K.}\ \bibnamefont
  {Gomes}}, \bibinfo {author} {\bibfnamefont {W.}~\bibnamefont {Mar}}, \bibinfo
  {author} {\bibfnamefont {W.}~\bibnamefont {Ko}}, \bibinfo {author}
  {\bibfnamefont {F.}~\bibnamefont {Guinea}}, \ and\ \bibinfo {author}
  {\bibfnamefont {H.~C.}\ \bibnamefont {Manoharan}},\ }\href
  {http://dx.doi.org/10.1038/nature10941} {\bibfield  {journal} {\bibinfo
  {journal} {Nature}\ }\textbf {\bibinfo {volume} {483}},\ \bibinfo {pages}
  {306} (\bibinfo {year} {2012})}\BibitemShut {NoStop}%
\bibitem [{\citenamefont {Rechtsman}\ \emph {et~al.}(2013)\citenamefont
  {Rechtsman}, \citenamefont {Zeuner}, \citenamefont {Tunnermann},
  \citenamefont {Nolte}, \citenamefont {Segev},\ and\ \citenamefont
  {Szameit}}]{Rechtsman2013}%
  \BibitemOpen
  \bibfield  {author} {\bibinfo {author} {\bibfnamefont {M.~C.}\ \bibnamefont
  {Rechtsman}}, \bibinfo {author} {\bibfnamefont {J.~M.}\ \bibnamefont
  {Zeuner}}, \bibinfo {author} {\bibfnamefont {A.}~\bibnamefont {Tunnermann}},
  \bibinfo {author} {\bibfnamefont {S.}~\bibnamefont {Nolte}}, \bibinfo
  {author} {\bibfnamefont {M.}~\bibnamefont {Segev}}, \ and\ \bibinfo {author}
  {\bibfnamefont {A.}~\bibnamefont {Szameit}},\ }\href
  {http://dx.doi.org/10.1038/nphoton.2012.302} {\bibfield  {journal} {\bibinfo
  {journal} {Nature Photon.}\ }\textbf {\bibinfo {volume} {7}},\ \bibinfo
  {pages} {153} (\bibinfo {year} {2013})}\BibitemShut {NoStop}%
\bibitem [{\citenamefont {Herbut}(2008)}]{Herbut2008}%
  \BibitemOpen
  \bibfield  {author} {\bibinfo {author} {\bibfnamefont {I.~F.}\ \bibnamefont
  {Herbut}},\ }\href {\doibase 10.1103/PhysRevB.78.205433} {\bibfield
  {journal} {\bibinfo  {journal} {Phys. Rev. B}\ }\textbf {\bibinfo {volume}
  {78}},\ \bibinfo {pages} {205433} (\bibinfo {year} {2008})}\BibitemShut
  {NoStop}%
\bibitem [{\citenamefont {Abanin}\ and\ \citenamefont
  {Pesin}(2012)}]{Abanin2012}%
  \BibitemOpen
  \bibfield  {author} {\bibinfo {author} {\bibfnamefont {D.~A.}\ \bibnamefont
  {Abanin}}\ and\ \bibinfo {author} {\bibfnamefont {D.~A.}\ \bibnamefont
  {Pesin}},\ }\href {\doibase 10.1103/PhysRevLett.109.066802} {\bibfield
  {journal} {\bibinfo  {journal} {Phys. Rev. Lett.}\ }\textbf {\bibinfo
  {volume} {109}},\ \bibinfo {pages} {066802} (\bibinfo {year}
  {2012})}\BibitemShut {NoStop}%
\bibitem [{\citenamefont {Gopalakrishnan}\ \emph {et~al.}(2012)\citenamefont
  {Gopalakrishnan}, \citenamefont {Ghaemi},\ and\ \citenamefont
  {Ryu}}]{Gopalakrishnan2012}%
  \BibitemOpen
  \bibfield  {author} {\bibinfo {author} {\bibfnamefont {S.}~\bibnamefont
  {Gopalakrishnan}}, \bibinfo {author} {\bibfnamefont {P.}~\bibnamefont
  {Ghaemi}}, \ and\ \bibinfo {author} {\bibfnamefont {S.}~\bibnamefont {Ryu}},\
  }\href {\doibase 10.1103/PhysRevB.86.081403} {\bibfield  {journal} {\bibinfo
  {journal} {Phys. Rev. B}\ }\textbf {\bibinfo {volume} {86}},\ \bibinfo
  {pages} {081403} (\bibinfo {year} {2012})}\BibitemShut {NoStop}%
\bibitem [{\citenamefont {Roy}\ and\ \citenamefont {Herbut}(2013)}]{Roy2013}%
  \BibitemOpen
  \bibfield  {author} {\bibinfo {author} {\bibfnamefont {B.}~\bibnamefont
  {Roy}}\ and\ \bibinfo {author} {\bibfnamefont {I.~F.}\ \bibnamefont
  {Herbut}},\ }\href {\doibase 10.1103/PhysRevB.88.045425} {\bibfield
  {journal} {\bibinfo  {journal} {Phys. Rev. B}\ }\textbf {\bibinfo {volume}
  {88}},\ \bibinfo {pages} {045425} (\bibinfo {year} {2013})}\BibitemShut
  {NoStop}%
\bibitem [{\citenamefont {Roy}\ \emph {et~al.}(2014{\natexlab{b}})\citenamefont
  {Roy}, \citenamefont {Assaad},\ and\ \citenamefont {Herbut}}]{Roy2014}%
  \BibitemOpen
  \bibfield  {author} {\bibinfo {author} {\bibfnamefont {B.}~\bibnamefont
  {Roy}}, \bibinfo {author} {\bibfnamefont {F.~F.}\ \bibnamefont {Assaad}}, \
  and\ \bibinfo {author} {\bibfnamefont {I.~F.}\ \bibnamefont {Herbut}},\
  }\href {\doibase 10.1103/PhysRevX.4.021042} {\bibfield  {journal} {\bibinfo
  {journal} {Phys. Rev. X}\ }\textbf {\bibinfo {volume} {4}},\ \bibinfo {pages}
  {021042} (\bibinfo {year} {2014}{\natexlab{b}})}\BibitemShut {NoStop}%
\bibitem [{\citenamefont {Roy}\ and\ \citenamefont {Sau}(2014)}]{Roy2014b}%
  \BibitemOpen
  \bibfield  {author} {\bibinfo {author} {\bibfnamefont {B.}~\bibnamefont
  {Roy}}\ and\ \bibinfo {author} {\bibfnamefont {J.~D.}\ \bibnamefont {Sau}},\
  }\href {\doibase 10.1103/PhysRevB.90.075427} {\bibfield  {journal} {\bibinfo
  {journal} {Phys. Rev. B}\ }\textbf {\bibinfo {volume} {90}},\ \bibinfo
  {pages} {075427} (\bibinfo {year} {2014})}\BibitemShut {NoStop}%
\bibitem [{\citenamefont {Soltan-Panahi}\ \emph {et~al.}(2011)\citenamefont
  {Soltan-Panahi}, \citenamefont {Struck}, \citenamefont {Hauke}, \citenamefont
  {Bick}, \citenamefont {Plenkers}, \citenamefont {Meineke}, \citenamefont
  {Becker}, \citenamefont {Windpassinger}, \citenamefont {Lewenstein},\ and\
  \citenamefont {Sengstock}}]{soltan2011multi}%
  \BibitemOpen
  \bibfield  {author} {\bibinfo {author} {\bibfnamefont {P.}~\bibnamefont
  {Soltan-Panahi}}, \bibinfo {author} {\bibfnamefont {J.}~\bibnamefont
  {Struck}}, \bibinfo {author} {\bibfnamefont {P.}~\bibnamefont {Hauke}},
  \bibinfo {author} {\bibfnamefont {A.}~\bibnamefont {Bick}}, \bibinfo {author}
  {\bibfnamefont {W.}~\bibnamefont {Plenkers}}, \bibinfo {author}
  {\bibfnamefont {G.}~\bibnamefont {Meineke}}, \bibinfo {author} {\bibfnamefont
  {C.}~\bibnamefont {Becker}}, \bibinfo {author} {\bibfnamefont
  {P.}~\bibnamefont {Windpassinger}}, \bibinfo {author} {\bibfnamefont
  {M.}~\bibnamefont {Lewenstein}}, \ and\ \bibinfo {author} {\bibfnamefont
  {K.}~\bibnamefont {Sengstock}},\ }\href@noop {} {\bibfield  {journal}
  {\bibinfo  {journal} {Nature Phys.}\ }\textbf {\bibinfo {volume} {7}},\
  \bibinfo {pages} {434} (\bibinfo {year} {2011})}\BibitemShut {NoStop}%
\bibitem [{\citenamefont {Duca}\ \emph {et~al.}(2015)\citenamefont {Duca},
  \citenamefont {Li}, \citenamefont {Reitter}, \citenamefont {Bloch},
  \citenamefont {Schleier-Smith},\ and\ \citenamefont {Schneider}}]{Duca2015}%
  \BibitemOpen
  \bibfield  {author} {\bibinfo {author} {\bibfnamefont {L.}~\bibnamefont
  {Duca}}, \bibinfo {author} {\bibfnamefont {T.}~\bibnamefont {Li}}, \bibinfo
  {author} {\bibfnamefont {M.}~\bibnamefont {Reitter}}, \bibinfo {author}
  {\bibfnamefont {I.}~\bibnamefont {Bloch}}, \bibinfo {author} {\bibfnamefont
  {M.}~\bibnamefont {Schleier-Smith}}, \ and\ \bibinfo {author} {\bibfnamefont
  {U.}~\bibnamefont {Schneider}},\ }\href {\doibase 10.1126/science.1259052}
  {\bibfield  {journal} {\bibinfo  {journal} {Science}\ }\textbf {\bibinfo
  {volume} {347}},\ \bibinfo {pages} {288} (\bibinfo {year}
  {2015})}\BibitemShut {NoStop}%
\bibitem [{\citenamefont {Alba}\ \emph {et~al.}(2013)\citenamefont {Alba},
  \citenamefont {Fernandez-Gonzalvo}, \citenamefont {Mur-Petit}, \citenamefont
  {Garcia-Ripoll},\ and\ \citenamefont {Pachos}}]{alba2013simulating}%
  \BibitemOpen
  \bibfield  {author} {\bibinfo {author} {\bibfnamefont {E.}~\bibnamefont
  {Alba}}, \bibinfo {author} {\bibfnamefont {X.}~\bibnamefont
  {Fernandez-Gonzalvo}}, \bibinfo {author} {\bibfnamefont {J.}~\bibnamefont
  {Mur-Petit}}, \bibinfo {author} {\bibfnamefont {J.~J.}\ \bibnamefont
  {Garcia-Ripoll}}, \ and\ \bibinfo {author} {\bibfnamefont {J.~K.}\
  \bibnamefont {Pachos}},\ }\href@noop {} {\bibfield  {journal} {\bibinfo
  {journal} {Annals of Physics}\ }\textbf {\bibinfo {volume} {328}},\ \bibinfo
  {pages} {64} (\bibinfo {year} {2013})}\BibitemShut {NoStop}%
\bibitem [{Note1()}]{Note1}%
  \BibitemOpen
  \bibinfo {note} {Since we are interested in the case $I_m\gtrsim 2 E_R$ we
  only need to consider nearest neighbor hopping ($t_\protect \text {nnn}=0.02
  t_\protect \text {nn}$ at $2 E_R$).}\BibitemShut {Stop}%
\bibitem [{\citenamefont {Suzuura}\ and\ \citenamefont
  {Ando}(2002)}]{Suzuura2002}%
  \BibitemOpen
  \bibfield  {author} {\bibinfo {author} {\bibfnamefont {H.}~\bibnamefont
  {Suzuura}}\ and\ \bibinfo {author} {\bibfnamefont {T.}~\bibnamefont {Ando}},\
  }\href {\doibase 10.1103/PhysRevB.65.235412} {\bibfield  {journal} {\bibinfo
  {journal} {Phys. Rev. B}\ }\textbf {\bibinfo {volume} {65}},\ \bibinfo
  {pages} {235412} (\bibinfo {year} {2002})}\BibitemShut {NoStop}%
\bibitem [{\citenamefont {Ma\~nes}(2007)}]{Manes2007}%
  \BibitemOpen
  \bibfield  {author} {\bibinfo {author} {\bibfnamefont {J.~L.}\ \bibnamefont
  {Ma\~nes}},\ }\href {\doibase 10.1103/PhysRevB.76.045430} {\bibfield
  {journal} {\bibinfo  {journal} {Phys. Rev. B}\ }\textbf {\bibinfo {volume}
  {76}},\ \bibinfo {pages} {045430} (\bibinfo {year} {2007})}\BibitemShut
  {NoStop}%
\bibitem [{\citenamefont {Walters}\ \emph {et~al.}(2013)\citenamefont
  {Walters}, \citenamefont {Cotugno}, \citenamefont {Johnson}, \citenamefont
  {Clark},\ and\ \citenamefont {Jaksch}}]{walters2013ab}%
  \BibitemOpen
  \bibfield  {author} {\bibinfo {author} {\bibfnamefont {R.}~\bibnamefont
  {Walters}}, \bibinfo {author} {\bibfnamefont {G.}~\bibnamefont {Cotugno}},
  \bibinfo {author} {\bibfnamefont {T.~H.}\ \bibnamefont {Johnson}}, \bibinfo
  {author} {\bibfnamefont {S.~R.}\ \bibnamefont {Clark}}, \ and\ \bibinfo
  {author} {\bibfnamefont {D.}~\bibnamefont {Jaksch}},\ }\href@noop {}
  {\bibfield  {journal} {\bibinfo  {journal} {Phys. Rev. A}\ }\textbf {\bibinfo
  {volume} {87}},\ \bibinfo {pages} {043613} (\bibinfo {year}
  {2013})}\BibitemShut {NoStop}%
\bibitem [{\citenamefont {Berestetskii}\ \emph {et~al.}(1971)\citenamefont
  {Berestetskii}, \citenamefont {Lifshitz},\ and\ \citenamefont
  {Pitaevskii}}]{Berestetskii1971}%
  \BibitemOpen
  \bibfield  {author} {\bibinfo {author} {\bibfnamefont {V.~B.}\ \bibnamefont
  {Berestetskii}}, \bibinfo {author} {\bibfnamefont {E.~M.}\ \bibnamefont
  {Lifshitz}}, \ and\ \bibinfo {author} {\bibfnamefont {L.~P.}\ \bibnamefont
  {Pitaevskii}},\ }\href@noop {} {\emph {\bibinfo {title} {Relativistic Quantum
  Theory}}}\ (\bibinfo  {publisher} {Pergamon, Oxford},\ \bibinfo {year}
  {1971})\BibitemShut {NoStop}%
\bibitem [{\citenamefont {Novoselov}\ \emph {et~al.}(2005)\citenamefont
  {Novoselov}, \citenamefont {Geim}, \citenamefont {Morozov}, \citenamefont
  {Jiang}, \citenamefont {Katsnelson}, \citenamefont {Grigorieva},
  \citenamefont {Dubonos},\ and\ \citenamefont {Firsov}}]{Novoselov2005}%
  \BibitemOpen
  \bibfield  {author} {\bibinfo {author} {\bibfnamefont {K.~S.}\ \bibnamefont
  {Novoselov}}, \bibinfo {author} {\bibfnamefont {A.~K.}\ \bibnamefont {Geim}},
  \bibinfo {author} {\bibfnamefont {S.~V.}\ \bibnamefont {Morozov}}, \bibinfo
  {author} {\bibfnamefont {D.}~\bibnamefont {Jiang}}, \bibinfo {author}
  {\bibfnamefont {M.~I.}\ \bibnamefont {Katsnelson}}, \bibinfo {author}
  {\bibfnamefont {I.~V.}\ \bibnamefont {Grigorieva}}, \bibinfo {author}
  {\bibfnamefont {S.~V.}\ \bibnamefont {Dubonos}}, \ and\ \bibinfo {author}
  {\bibfnamefont {A.~A.}\ \bibnamefont {Firsov}},\ }\href@noop {} {\bibfield
  {journal} {\bibinfo  {journal} {Nature}\ }\textbf {\bibinfo {volume} {438}},\
  \bibinfo {pages} {197} (\bibinfo {year} {2005})}\BibitemShut {NoStop}%
\bibitem [{\citenamefont {Greif}\ \emph {et~al.}(2013)\citenamefont {Greif},
  \citenamefont {Uehlinger}, \citenamefont {Jotzu}, \citenamefont {Tarruell},\
  and\ \citenamefont {Esslinger}}]{Greif2013}%
  \BibitemOpen
  \bibfield  {author} {\bibinfo {author} {\bibfnamefont {D.}~\bibnamefont
  {Greif}}, \bibinfo {author} {\bibfnamefont {T.}~\bibnamefont {Uehlinger}},
  \bibinfo {author} {\bibfnamefont {G.}~\bibnamefont {Jotzu}}, \bibinfo
  {author} {\bibfnamefont {L.}~\bibnamefont {Tarruell}}, \ and\ \bibinfo
  {author} {\bibfnamefont {T.}~\bibnamefont {Esslinger}},\ }\href {\doibase
  10.1126/science.1236362} {\bibfield  {journal} {\bibinfo  {journal}
  {Science}\ }\textbf {\bibinfo {volume} {340}},\ \bibinfo {pages} {1307}
  (\bibinfo {year} {2013})}\BibitemShut {NoStop}%
\bibitem [{\citenamefont {Hart}\ \emph {et~al.}(2015)\citenamefont {Hart},
  \citenamefont {Duarte}, \citenamefont {Yang}, \citenamefont {Liu},
  \citenamefont {Paiva}, \citenamefont {Khatami}, \citenamefont {Scalettar},
  \citenamefont {Trivedi}, \citenamefont {Huse},\ and\ \citenamefont
  {Hulet}}]{Hart2015}%
  \BibitemOpen
  \bibfield  {author} {\bibinfo {author} {\bibfnamefont {R.~A.}\ \bibnamefont
  {Hart}}, \bibinfo {author} {\bibfnamefont {P.~M.}\ \bibnamefont {Duarte}},
  \bibinfo {author} {\bibfnamefont {T.-L.}\ \bibnamefont {Yang}}, \bibinfo
  {author} {\bibfnamefont {X.}~\bibnamefont {Liu}}, \bibinfo {author}
  {\bibfnamefont {T.}~\bibnamefont {Paiva}}, \bibinfo {author} {\bibfnamefont
  {E.}~\bibnamefont {Khatami}}, \bibinfo {author} {\bibfnamefont {R.~T.}\
  \bibnamefont {Scalettar}}, \bibinfo {author} {\bibfnamefont {N.}~\bibnamefont
  {Trivedi}}, \bibinfo {author} {\bibfnamefont {D.~A.}\ \bibnamefont {Huse}}, \
  and\ \bibinfo {author} {\bibfnamefont {R.~G.}\ \bibnamefont {Hulet}},\ }\href
  {http://dx.doi.org/10.1038/nature14223} {\bibfield  {journal} {\bibinfo
  {journal} {Nature}\ }\textbf {\bibinfo {volume} {519}},\ \bibinfo {pages}
  {211} (\bibinfo {year} {2015})}\BibitemShut {NoStop}%
\bibitem [{\citenamefont {Ernst}\ \emph {et~al.}(2010)\citenamefont {Ernst},
  \citenamefont {Gotze}, \citenamefont {Krauser}, \citenamefont {Pyka},
  \citenamefont {Luhmann}, \citenamefont {Pfannkuche},\ and\ \citenamefont
  {Sengstock}}]{Ernst2010}%
  \BibitemOpen
  \bibfield  {author} {\bibinfo {author} {\bibfnamefont {P.~T.}\ \bibnamefont
  {Ernst}}, \bibinfo {author} {\bibfnamefont {S.}~\bibnamefont {Gotze}},
  \bibinfo {author} {\bibfnamefont {J.~S.}\ \bibnamefont {Krauser}}, \bibinfo
  {author} {\bibfnamefont {K.}~\bibnamefont {Pyka}}, \bibinfo {author}
  {\bibfnamefont {D.-S.}\ \bibnamefont {Luhmann}}, \bibinfo {author}
  {\bibfnamefont {D.}~\bibnamefont {Pfannkuche}}, \ and\ \bibinfo {author}
  {\bibfnamefont {K.}~\bibnamefont {Sengstock}},\ }\href@noop {} {\bibfield
  {journal} {\bibinfo  {journal} {Nature Phys.}\ }\textbf {\bibinfo {volume}
  {6}},\ \bibinfo {pages} {56} (\bibinfo {year} {2010})}\BibitemShut {NoStop}%
\bibitem [{\citenamefont {Kling}\ \emph {et~al.}(2010)\citenamefont {Kling},
  \citenamefont {Salger}, \citenamefont {Grossert},\ and\ \citenamefont
  {Weitz}}]{Kling2010}%
  \BibitemOpen
  \bibfield  {author} {\bibinfo {author} {\bibfnamefont {S.}~\bibnamefont
  {Kling}}, \bibinfo {author} {\bibfnamefont {T.}~\bibnamefont {Salger}},
  \bibinfo {author} {\bibfnamefont {C.}~\bibnamefont {Grossert}}, \ and\
  \bibinfo {author} {\bibfnamefont {M.}~\bibnamefont {Weitz}},\ }\href
  {\doibase 10.1103/PhysRevLett.105.215301} {\bibfield  {journal} {\bibinfo
  {journal} {Phys. Rev. Lett.}\ }\textbf {\bibinfo {volume} {105}},\ \bibinfo
  {pages} {215301} (\bibinfo {year} {2010})}\BibitemShut {NoStop}%
\bibitem [{\citenamefont {Macri}\ and\ \citenamefont
  {Pohl}(2014)}]{macri2014rydberg}%
  \BibitemOpen
  \bibfield  {author} {\bibinfo {author} {\bibfnamefont {T.}~\bibnamefont
  {Macri}}\ and\ \bibinfo {author} {\bibfnamefont {T.}~\bibnamefont {Pohl}},\
  }\href@noop {} {\bibfield  {journal} {\bibinfo  {journal} {Phys. Rev. A}\
  }\textbf {\bibinfo {volume} {89}},\ \bibinfo {pages} {011402} (\bibinfo
  {year} {2014})}\BibitemShut {NoStop}%
\bibitem [{\citenamefont {De~Paz}\ \emph {et~al.}(2013)\citenamefont {De~Paz},
  \citenamefont {Sharma}, \citenamefont {Chotia}, \citenamefont {Marechal},
  \citenamefont {Huckans}, \citenamefont {Pedri}, \citenamefont {Santos},
  \citenamefont {Gorceix}, \citenamefont {Vernac},\ and\ \citenamefont
  {Laburthe-Tolra}}]{de2013nonequilibrium}%
  \BibitemOpen
  \bibfield  {author} {\bibinfo {author} {\bibfnamefont {A.}~\bibnamefont
  {De~Paz}}, \bibinfo {author} {\bibfnamefont {A.}~\bibnamefont {Sharma}},
  \bibinfo {author} {\bibfnamefont {A.}~\bibnamefont {Chotia}}, \bibinfo
  {author} {\bibfnamefont {E.}~\bibnamefont {Marechal}}, \bibinfo {author}
  {\bibfnamefont {J.}~\bibnamefont {Huckans}}, \bibinfo {author} {\bibfnamefont
  {P.}~\bibnamefont {Pedri}}, \bibinfo {author} {\bibfnamefont
  {L.}~\bibnamefont {Santos}}, \bibinfo {author} {\bibfnamefont
  {O.}~\bibnamefont {Gorceix}}, \bibinfo {author} {\bibfnamefont
  {L.}~\bibnamefont {Vernac}}, \ and\ \bibinfo {author} {\bibfnamefont
  {B.}~\bibnamefont {Laburthe-Tolra}},\ }\href@noop {} {\bibfield  {journal}
  {\bibinfo  {journal} {Phys. Rev. Lett.}\ }\textbf {\bibinfo {volume} {111}},\
  \bibinfo {pages} {185305} (\bibinfo {year} {2013})}\BibitemShut {NoStop}%
\bibitem [{\citenamefont {Hazzard}\ \emph {et~al.}(2014)\citenamefont
  {Hazzard}, \citenamefont {Gadway}, \citenamefont {Foss-Feig}, \citenamefont
  {Yan}, \citenamefont {Moses}, \citenamefont {Covey}, \citenamefont {Yao},
  \citenamefont {Lukin}, \citenamefont {Ye}, \citenamefont {Jin},\ and\
  \citenamefont {Rey}}]{hazzard2014many}%
  \BibitemOpen
  \bibfield  {author} {\bibinfo {author} {\bibfnamefont {K.~R.}\ \bibnamefont
  {Hazzard}}, \bibinfo {author} {\bibfnamefont {B.}~\bibnamefont {Gadway}},
  \bibinfo {author} {\bibfnamefont {M.}~\bibnamefont {Foss-Feig}}, \bibinfo
  {author} {\bibfnamefont {B.}~\bibnamefont {Yan}}, \bibinfo {author}
  {\bibfnamefont {S.~A.}\ \bibnamefont {Moses}}, \bibinfo {author}
  {\bibfnamefont {J.~P.}\ \bibnamefont {Covey}}, \bibinfo {author}
  {\bibfnamefont {N.~Y.}\ \bibnamefont {Yao}}, \bibinfo {author} {\bibfnamefont
  {M.~D.}\ \bibnamefont {Lukin}}, \bibinfo {author} {\bibfnamefont
  {J.}~\bibnamefont {Ye}}, \bibinfo {author} {\bibfnamefont {D.~S.}\
  \bibnamefont {Jin}}, \ and\ \bibinfo {author} {\bibfnamefont {A.~M.}\
  \bibnamefont {Rey}},\ }\href@noop {} {\bibfield  {journal} {\bibinfo
  {journal} {Phys. Rev. Lett.}\ }\textbf {\bibinfo {volume} {113}},\ \bibinfo
  {pages} {195302} (\bibinfo {year} {2014})}\BibitemShut {NoStop}%
\end{thebibliography}%

\end{document}